# A NOVEL 2.5D APPROACH FOR INTERFACING WITH WEB APPLICATIONS

*Saurabh Sarkar*
**Tata Research Development and Design Centre, Pune**
`sarkarsaurabh.27@gmail.com`

*Abstract* – **Web applications need better user interface to be interactive and attractive. I present a new approach/concept of dimensional enhancement – 2.5D "a 2D display of a virtual 3D environment", which can be implemented in social networking sites and further in other system applications.**

## I. INTRODUCTION

The **user interface** [1], in the industrial design field of human–machine interaction, is the space where interaction between humans and machines occurs. The goal of interaction between a human and a machine at the user interface is effective operation and control of the machine, and feedback from the machine which aids the operator in making operational decisions.
Abbreviated *UI* [2], the junction between a user and a computer program. UI is one of the most important parts of any program because it determines how easily you can make the program do what you want.
The User Interface (UI) of an application (Web or System) can make it successful in terms of business and efficiency. The more interactive and attractive the UI is, more the user uses it with ease. Many technological innovations rely upon User Interface Design to elevate their technical complexity to a usable product. Technology alone may not win user acceptance and subsequent marketability [3]. The UI of a system application eliminates the need for the user to learn how the system interacts with the hardware; it just gives the user a platform to get what the user desires to. In case of a web application, UI is the part which makes it successful. Let's consider Social Networking sites.

This paper details 2.5 D, the new approach/concept of dimensional enhancement which can be used to make social networking sites [4] more successful by use of better and innovative UI. How we can achieve it with an enhancement of a dimensional perspective? Social networking sites are already pretty popular. So how do you improve on the internet's hottest web offerings?

## II. BACKGROUND

In the knowledge society, in addition to technical skills and access to information technologies, it is becoming increasingly important for people to have diversified and supportive social connections. Many management practitioners have also confirmed how social networking has changed business [5]. Social networking morphed from a personal communications tool for young people into a new vehicle that business leaders are using to transform communications with their employees and customers, as it shifts from one-way transmission of information to two-way interaction. The reviews of various social networks can be viewed [6].

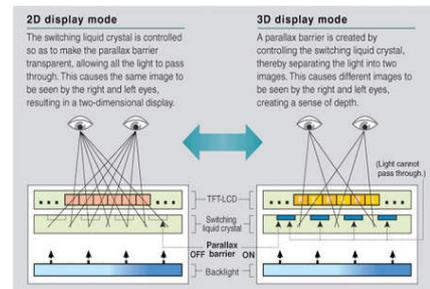

**Fig. 1**. Description of 2D and 3D display modes

With the rise of virtual environments, augmented reality, large- screen display systems, and three-dimensional (3D) applications of all sorts on the desktop, a new trend in human -computer interaction (HCI) research began to emerge. 3D technology attracts masses and if this technology can be applied efficiently in applications [7], it would make the applications attractive. Although principles gleaned from years of experience in designing UIs for desktop computers still applied, they weren't sufficient to address the unique needs of these systems where interaction took place in a 3D spatial context, with multiple degrees-of-freedom, due to the lack the hardware requirements.

*Fig. 1* shows the working of both the 2D and 3D display modes. Implementing any new technologies has some dependencies and demerits. In case of 3D technology, there is a requirement of hardware i.e. dependency on input and output devices [8]. These include spatial input devices such as trackers, 3-D pointing devices, and whole-hand devices allowing gestural input. Three-dimensional, multi-sensory output technologies - such as stereoscopic projection displays, head-mounted displays (HMDs), spatial audio systems, and haptic devices. Over that, in the current state of art various real 3D displays are introduced like displays without the use of 3D glasses. But this is achieved by the use of special kind of LCDs as the one introduced by *Nintendo*. But our aim is to provide users a way to achieve the Virtual Reality Software and Technology in web applications (Social Networking sites in our case). Our goals are to describe some of the major interaction techniques and interface components that can be implemented in web applications.

Web3D is a booming technology now-a-days. Our concept mainly deals with how we can make the user-experience better to improve technical aspects as well as business for any organization. We are mainly focusing on how this experience can be accessed by a comparatively lower system configuration and the current web technology only. The innovation deals with the designing part and how we can provide the interface within any web browser.

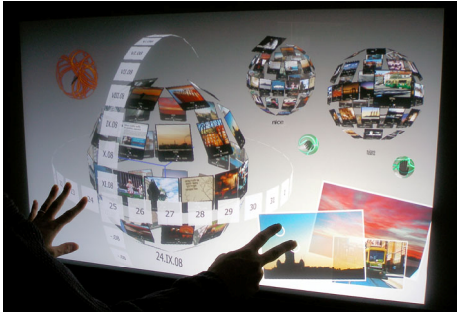

**Fig. 2.** A UI which can be used in social networking sites for sharing and interacting on click/touch © CityWall, Helsinki

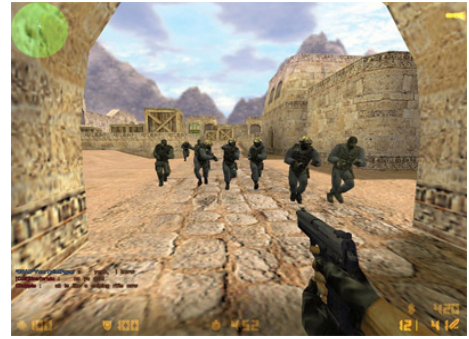

**Fig. 3**. A snapshot of 2.5D (LAN game – Counter Strike © Valve Software)

Many client applications needs to be installed on the system accessing the configuration, but if we are offering this service on the web-browsers, through a web-site, there won't be any direct access to the system configurations, so it needs to be dynamic and robust. Moreover, we are focusing on the limitations like Internet connections capable of smoothly rendering high quality content like in massively multiplayer online role-playing game, performing poorly, resulting in low frame rates and unresponsive controls on even minimal graphical configurations and low bandwidth. So our main focus is on the user experience through an attractive 3D like UI and how we can make this possible with maximum data transfer though low bandwidth in real time.

III. PROPOSED THEORY

To achieve a view or interface that gives a 3D feel but does not have any dependency for that of input or output devices. To give user a UI that can help them to achieve the functionality they require. We introduce a dimensional perspective that is not exactly 3D but gives a feel of the user's presence in the virtual world. This area of research gradually came to be known as *2.5D User Interfaces (2.5D UIs), or 2.5D Interaction*. I would not say that it is a step back to the 3D technology which already exists but a proposed dimension which can be easily achieved by everyone without any dependencies as of 3D and also gives a feel of 3D.

We define **2.5D interaction** as "the connecting bridge between 2D and 3D where both dimensions are taken together and presented in an economical and feasible manner that can be accessed by every type of user".

There are some platforms already present where similar types of technology have already been developed and are used, by some users in different domains but are not used as a UI for any application. I would like to implement these technologies by enhancing them according to our problem statement.

Web technologies provide the basis to share and access digital information globally and in real time but they have also established the Web as a ubiquitous application platform. We are trying to extend core Web technologies to support interactive 3D content. Current technologies like HTML5 and XHTML with JavaScript can help achieve somewhat of the required results with the help of some programming languages.

XML3D enables portable cross-platform authoring, distribution, and rendering of and interaction with 3D data. XML3D attempts to achieve maximum compatibility with both HTML5 and XHTML. Many of XML3D's facilities are modeled directly after HTML and SVG, including its use of CSS, its approach to event handling, and its approach to the Document Object Model. The aim is to create a web standard that is easy to learn by web designers and web programmers without deep knowledge about 3D programming. We are currently working on specific browser frameworks. Our implementation plan is to work with implicit installation of data on the system like virtual default and custom maps, etc. and just with the help of plug-ins allowing user to access the application with minimum data transfer each time and in real time. I cannot exactly say that it is easy to implement, because the data transfer depends on the network. So our main focus of design and data transfer is trying to deal with that.

IV. CASE STUDIES OF NEW DIRECTIONS

Let us consider an example of implementation of 2.5D - a LAN game that connects users globally and helps them to interact and play. The game is *Counter Strike* which helps users to feel their presence with others in that virtual world and helps them share their lives. Games like Counter Strike are as popular as Microsoft products which help users to be together in a virtual world because of the interface. More games like *Age of Wars* provide a platform to chat and interact with the other users on network. Or a Virtual World likes *Second Life.* Our approach aims to give users such an experience using web-browsers, so that user would not feel alone in this digital world.

It's not a plug-in which can be applied to other websites and transform them to 3D i.e. what *Exit Reality* does, but it's a website which will provide you a virtual world not so different from our real world.

Fig. 3 shows us a snapshot of the LAN game, where a person can view other users around them. This approach can be used and to merge a 2.5DUI similar to this in social networking to give users **the 2D interactivity in a virtual 3D world** as also mentioned in Ref. [8]. It may be implemented using some script languages which can be easily transferred through network and be provided to the users.

## V. CONCLUSION AND FUTURE WORK

In this paper, I have proposed an approach which can make social networking (communication among the world) more interactive and attractive. I have introduced a new concept of *2.5D UI* which is **a 2D display of a virtual 3D environment** which can be used in these types of web applications. But providing such technology over network has some dependency of network speed (bandwidth). If we implement this on any web application, the user must access it on a high bandwidth network so that all data can be transferred easily, which can be discussed further.

This approach can further be used in any system applications also, to make the interface between the user and the hardware much more interesting and easy. So that the user can feel the application as a physical system of his world.

This is an era of rich user experience so there will be insatiable demand for this hence movement from 2D to 2.5D is more of when rather than will but from 2.5D to 3D might be restricted because of hardware requirements as mentioned in the earlier interaction that the next billion internet users will access the net from their mobile phones. As a majority of the system users are utility system users not consumer system users, so it is improbable that the industrial users would enhances their systems for accessing 3D experience, so to give them a 3D experience on 2D hardware support, this 2.5D would be the best technology. Will 3d technology come to mobiles, probably yes, but in the near future seems unlikely. Cost is a big deterrent, at least may not be a big hit in the third world countries where the number of internet users is increasing as compared to the ones in developing countries where it is stagnating.

If our study goes well, and we are able to overcome all the limitations faced earlier, which I already mentioned in my earlier mail, we are planning to start with a small application for example in facebook, that would enable all the users to experience the 2.5D UI, so that with a confirmation of its performance we can go to a fully fledged web application for all. Later in system applications, as well.